# The Non-Relativistic Limit of Keldysh Spinors

A. Jourjine[1]

FG CTP, Augsburger Str.82, 01277 Dresden, Germany

Abstract

Keldysh spinors obey Dirac equation, but have the negative of the Dirac action and Hamiltonian. In an example of the *U(1) EM* coupling, we show that, despite the sign changes, they have a well-defined non-relativistic limit resulting in quantum mechanics with the positive-definite Pauli Hamiltonian. When non-relativistic Dirac and Keldysh fields are brought to interact, we observe curious decoupling of the two fields in mass-like and vector couplings.

## 1. Introduction

Spinors that obey Dirac equation but whose action is the negative of the Dirac action. appear in widely different areas of Physics, such as the Keldysh non-equilibrium QFT [1, 2], in theories of flavor mixing that use no flavons [3, 4], and in research of the black hole information paradox [5]. To distinguish these particles from the positive action Dirac particles we call them the Keldysh spinors.

Because Keldysh spinors have Hamiltonian that is the negative of the Dirac Hamiltonian, two questions arise. These are can one perform the second quantization of such fields to obtain a positive-definite Hamiltonian operator and is the corresponding QFT has the correct non-relativistic limit that describes time evolution by the Schrödinger equation with positive definite Pauli Hamiltonian?

The first question was answered in the positive in earlier publications [6, 7]. In this Note we also answer in the positive the second question. Despite apparent negativity, the non-relativistic limit of Keldysh spinors interacting with classical EM field is described by the Schrödinger equation. The only difference between the two cases is the time dependence of the positive energy solutions in the eigenvalue problem. This is derived in Section 2. Section 3 is the Summary

---

[1] jourjine@pks.mpg.de

## 2. Non-Relativistic Limit of Keldysh Spinors

The standard Dirac and Keldysh spinor actions, interacting with electromagnetic field with $U(1)$ vector potential $A_\mu$, are given by

$$S_D = +\int d^4x\, \bar\psi\, (i\partial - eA - m)\psi, \qquad (1)$$

$$S_K = -\int d^4x\, \bar\psi\, (i\partial - eA - m)\psi. \qquad (2)$$

As a result, the respective canonical momenta and Hamiltonians also differ by a sign $p_K = -i\psi^+ = -p_D$, $H_K = -H_D$. Obviously, in both cases equations of motion are the same $(i\partial - eA - m)\psi = 0$. If the classical version of $H_D$ were positive definite, the reversal of the sign would have had fatal consequences and the quantized theory would have only negative energy states. This is certainly true for bosonic theories.

In the fermionic case on the classical level $H_D$ is indefinite and Dirac equation has both positive and negative energy solutions. Obviously the same applies to Keldysh spinors. To make $H_D$ positive definite one has to use additional information about anti-commuting of the fields and specific rule of assignment of the creation and annihilation $(c-a)$ operators. As has been shown in [6], by swapping the standard $(c-a)$ operators one can make the quantum $H_K$ positive definite as well. In this note we will show that Keldysh spinors also have a well-defined non-relativistic limit with the same positive definite Pauli eigenvalue problem Hamiltonian as the Dirac spinors. This makes Dirac and Keldysh spinors indistinguishable on the quantum-mechanical level.

The derivation of the limit is similar to the Dirac case. The only difference arises when we need to define positive energy solutions of the Dirac equation. Obviously, for Keldysh case these are the negative energy solutions of the Dirac case. For convenience we derive the limit for both cases in parallel.

The non-relativistic limit of the Dirac equation can be obtained using the Dirac $\gamma$–matrix representation, where $\gamma^0 = \begin{pmatrix} I & 0 \\ 0 & -I \end{pmatrix}$, $\alpha_k \equiv \gamma^0 \gamma^k = \begin{pmatrix} 0 & \sigma^k \\ \sigma^k & 0 \end{pmatrix}$. In the two component notation $\psi = \begin{pmatrix} \varphi \\ \chi \end{pmatrix}$ the Dirac equation becomes

$$i\partial \varphi/\partial t = (\sigma \cdot \pi)\chi + eA^0 \varphi + m\varphi,$$
$$i\partial \chi/\partial t = (\sigma \cdot \pi)\varphi + eA^0 \chi - m\chi, \qquad (3)$$

where $\pi = -\alpha_k(i\partial_k + eA_k)$. The next step is to factor to factor out from $\varphi, \chi$ the fast vacuum phase. Here Dirac and Keldysh cases begin to differ, because we want to ensure that our quantum mechanical system has positive energy states for the eigenvalue problem. For Dirac case such positive energy phase is $\exp(-imt)$, while for Keldysh case it is $\exp(+imt)$. Accordingly, we obtain for $\varphi_D = e^{-imt}\phi_D, \chi_D = e^{-imt}\xi_D, \varphi_K = e^{+imt}\phi_K, \chi_K = e^{+imt}\xi_K$

$$i\partial\phi_D/\partial t = (\sigma\cdot\pi)\xi_D + eA^0\phi_D,$$
$$i\partial\xi_D/\partial t = (\sigma\cdot\pi)\phi_D + eA^0\xi_D - 2m\xi_D, \qquad (4)$$

$$i\partial\phi_K/\partial t = (\sigma\cdot\pi)\xi_K + eA^0\phi_K + 2m\phi_K,$$
$$i\partial\xi_K/\partial t = (\sigma\cdot\pi)\phi_K + eA^0\xi_K. \qquad (5)$$

Assuming $|eA^0| << m$ and slow varying $\xi_D, \phi_K$, the second equation in (4) and the first equation in (5) can be solved approximately as $\xi_D = (1/2m)(\sigma\cdot\pi)\phi_D$ and $\phi_K = -(1/2m)(\sigma\cdot\pi)\xi_K$. The remaining equations become

$$i\partial\phi_D/\partial t = +\left[\frac{1}{2m}(\sigma\cdot\pi)^2\phi_D + eA^0\phi_D\right], \qquad (6)$$

$$i\partial\xi_K/\partial t = -\left[\frac{1}{2m}(\sigma\cdot\pi)^2\xi_K + (-e)A^0\xi_K\right]. \qquad (7)$$

Despite the appearance, the eigenvalue equation for (7) is the same as for (6) up to the charge sign. This is because the positive energy eigenvalue problem solutions for the two equations have different time dependence $\exp(\mp iE_k t)$. After reduction of $(\sigma\cdot\pi)^2$ we see that both equations are the well-known Schrödinger equations with the corresponding Pauli Hamiltonians

$$H_{D,K} = \frac{1}{2m}(i\partial_k \pm eA_k)^2 \mp \frac{e}{2m}(\sigma_k B_k) \pm eA^0, \qquad (8)$$

where the magnetic field strength $B_k = \varepsilon_{klm}\partial_l A_m$.

## 3. Discussion

In summary, the non-relativistic limit of both Dirac and Keldysh spinors with positive energy states leads to the Pauli equations on the wavefunctions differing only in the sign of the electromagnetic coupling.

Pauli Hamiltonian in (8) is only the first term of the expansion of the full relativistic

Hamiltonian in powers of inverse mass that separates the upper and the lower components of the spinor field. The expansion is obtained by the well-known Foldy-Wouthuysen transformation. Obviously, this transformation produces the same power series for both Dirac and Keldysh spinors. This makes Dirac and Keldysh spinors indistinguishable in local low energy experiments.

Although when interacting among themselves the two types of spinors are indistinguishable, curious effects arise when the two types are brought together. To see this, let us return to the four component spinor notation. In the non-relativistic limit we have

$$\psi_D = e^{-imt}\begin{pmatrix}\phi_D \\ 0\end{pmatrix}, \qquad \psi_K = e^{+imt}\begin{pmatrix}0 \\ \xi_K\end{pmatrix}. \tag{9}$$

The first curiosity is that the scalar coupling vanishes, $\bar{\psi}_D \psi_K = 0$ but $\bar{\psi}_D \gamma^5 \psi_K \neq 0$. The second one appears in the $U(1)$ vector coupling with vector potential $A_\mu$, namely $A_0 \bar{\psi}_D \gamma^0 \psi_K = 0$ but $A_0 \bar{\psi}_D \gamma^0 \gamma^5 \psi_K \neq 0$. This means that static electric field cannot couple the two spinors. The remaining component of the vector coupling term is $A_k \bar{\psi}_D \gamma^k \psi_K$. Static $A_k$ produces only magnetic field, which cannot transfer energy, which means that there is no energy transfer between the two fields, but elastic scattering is allowed. Combining vector and axial coupling to form left-handed coupling results in another curious formula

$$A_\mu \bar{\psi}_D \sigma^\mu (1-\gamma^5) \psi_K = A_\mu \phi_D^* \sigma^\mu (1-\gamma^5) \xi_K,$$

where $\sigma^\mu = (1, \sigma^k)$.